\documentclass[conference]{IEEEtran}
\IEEEoverridecommandlockouts

\usepackage{cite}
\usepackage{amsmath,amssymb,amsfonts}
\usepackage{graphicx}
\usepackage{textcomp}
\usepackage{xcolor}
\usepackage{subfiles}
\usepackage{multirow}
\usepackage{float}
\usepackage{stfloats}
\usepackage[table,xcdraw]{xcolor}
\usepackage{amsthm}
\usepackage{mathtools}
\usepackage{booktabs}
\usepackage{multirow}
\usepackage{comment}
\usepackage{array}
\usepackage{siunitx}
\usepackage{makecell}
\usepackage[utf8]{inputenc}
\usepackage{algorithm}
\usepackage{algpseudocode}
\theoremstyle{definition}

\newtheorem{theorem}{Theorem}
\graphicspath{ {./figures/} }
\def\BibTeX{{\rm B\kern-.05em{\sc i\kern-.025em b}\kern-.08em
    T\kern-.1667em\lower.7ex\hbox{E}\kern-.125emX}}

\begin{document}
\newcommand\Mycomb[2][^n]{\prescript{#1\mkern-0.5mu}{}C_{#2}}

\title{Inverter Redistribution through Self-Dual and Self-Anti-Dual Function Transformation}

\author{\IEEEauthorblockN{Jingren Wang}
\IEEEauthorblockA{\textit{MICS, HKUST(GZ)}\\
Guangzhou, China}
\and
\IEEEauthorblockN{Guangyu Hu}
\IEEEauthorblockA{\textit{IIP-MICS, HKUST}\\
Hong Kong, China}
\and
\IEEEauthorblockN{Shiju Lin}
\IEEEauthorblockA{\textit{MICS, HKUST(GZ)}\\
Guangzhou, China}
\and
\IEEEauthorblockN{Hongce Zhang}
\IEEEauthorblockA{\textit{MICS, HKUST(GZ)}\\
Guangzhou, China}
}

\maketitle

\begin{abstract}
And-Inverter Graph (AIG)-based logic synthesis has been a cornerstone of digital design automation for several decades. While numerous optimization techniques have been developed for both technology-independent and technology-dependent synthesis stages, existing technology mapping approaches predominantly employ graph-covering strategies directly on AIG representations without adequately addressing complemented edge distribution. Neglecting inverters creates a significant disconnect: complemented edges are systematically overlooked in technology-independent cost functions, yet they abruptly become critical during technology-dependent mapping. In this work, we introduce a delay-driven pre-processing stage that operates prior to technology mapping, designed to strategically redistribute complemented edges and mitigate the inverter-induced costs on critical paths. Experimental validation demonstrates that our delay-targeted methodology not only preserves original delay characteristics but also enables performance improvements. Notably, arithmetic logic in the EPFL combinational benchmark exhibits particular sensitivity to this approach, with our method achieving an average delay reduction of 0.49\% and a maximum improvement of 3.86\% on the case \textit{sqrt}.
\end{abstract}

\begin{IEEEkeywords}
And-Inverter Graph, Inverters, Logic Synthesis, Self-Dual, Self-Anti-Dual, Technology Mapping
\end{IEEEkeywords}

\section{Introduction}\label{INTRO}
A persistent disconnect remains between technology-independent and technology-dependent optimization phases in contemporary logic synthesis methodologies. Notably, inverters receive minimal attention in SOTA~(state-of-the-art) combinational technology-independent techniques, particularly those employing And-Inverter Graph (AIG) representations. The AIG data structure conceptualizes inverters as complemented edge attributes, a design choice that systematically excludes them from conventional cost functions such as logic level and node count~\cite{Orchestrate}.

This abstraction, while convenient, introduces nontrivial challenges. During technology-independent optimization, a mixture of structural rewriting and algebraic transformations is applied. These transformations are primarily guided by local equivalence and do not explicitly account for inverter placement. As a result, the distribution of complemented edges evolves in an essentially uncontrolled manner, appearing almost random from a global perspective. Consequently, when the flow proceeds to technology mapping, inverter placement suddenly becomes a critical factor affecting delay and area, revealing a clear mismatch between earlier optimization objectives and downstream requirements.

This discontinuity motivates the need for inverter-aware optimization at the technology-independent stage. Without such consideration, intermediate networks may contain inverters on critical paths that could otherwise be relocated to non-critical regions or eliminated altogether. However, existing synthesis flows provide limited mechanisms to explicitly control inverter distribution prior to mapping.
In delay-driven mapping scenarios~\cite{ExactDSyn,speedup}, this limitation becomes particularly pronounced, as inverter placement directly impacts the achievable timing quality.  Therefore, we propose a pre-processing step that reshapes inverter distribution before mapping to improve the quality of implementations. Since technology-independent optimization is inherently heuristic, inverter redistribution must also be designed in a heuristic yet goal-oriented manner, balancing trade-offs such as area overhead, delay degradation, and potential hazards~\cite{Hazard}. The strategy should align with the intended optimization objectives of the designer.

\begin{figure*}[h]
    \centering
    \includegraphics[scale=0.1114]{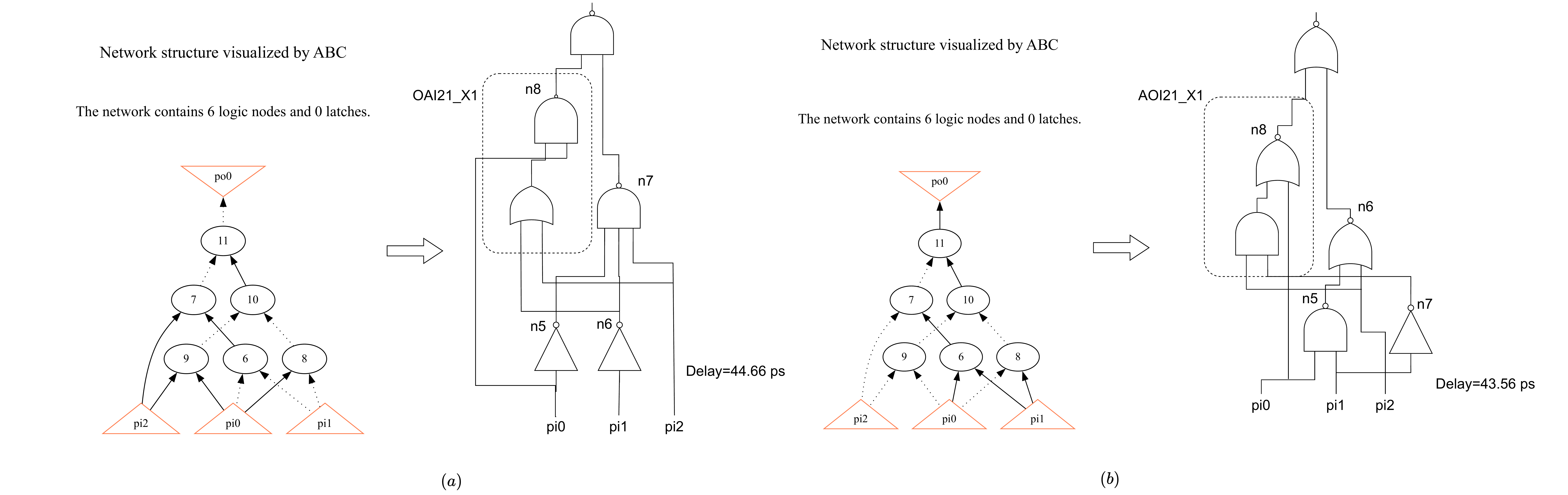}
    \caption{
        Both subfigures illustrate the technology mapping process from AIG to standard cells for an identical Boolean function. Dashed lines denote complemented (inverted) edges. Subfigure $(a)$ depicts the baseline mapping without inverter redistribution, while $(b)$ applies our proposed method by identifying a self-dual subfunction and redistributing its complemented edges, resulting in better delay performance.
    }
    \label{fig:MREIntro}
\end{figure*}

In this work, we propose a delay-oriented inverter redistribution technique applied as a preprocessing step prior to technology mapping, leveraging the self-dual and self-anti-dual functions (which will be explained in Section~\ref{SDSAD}).
We view this as an incremental yet practical approach to improving mapping quality. Figure~\ref{fig:MREIntro} presents a minimal reproducible example (MRE)~\cite{10SimpleRules} illustrating the impact of our method, where a self-dual subfunction is identified in the AIG, enabling the complementation of all cut inputs and the root fanout, meanwhile reducing inverters on the critical path.

Our primary contributions are as follows:
\begin{itemize}
    \item First, we introduce a novel inverter redistribution methodology specifically utilizing the properties of self-dual and self-anti-dual Boolean functions.
    \item Second, we implement this transformation within the AIG framework while preserving the original AND node structure, thereby maintaining the effects of prior optimization efforts in terms of node count and logic level.
\end{itemize}

\section{Background}\label{BG}
\subsection{Self-Dual/Self-Anti-Dual}\label{SDSAD}

Self-duality and self-anti-duality represent fundamental symmetry properties in Boolean algebra~\cite{STLS}. These properties exhibit unique characteristics that enable useful structural transformations in logic networks.
\begin{align}\label{FSD}
f(x_1, x_2, x_3, \cdots, x_n) = \bar f(\bar x_1, \bar x_2, \bar x_3, \cdots, \bar x_n)
\end{align}
Equation~\ref{FSD} formalizes the self-dual property: a function remains invariant under simultaneous complementation of all inputs and the function output. This symmetry enables powerful transformation opportunities.
\begin{align}\label{FSAD}
f(x_1, x_2, x_3, \cdots, x_n) = f(\bar x_1, \bar x_2, \bar x_3, \cdots, \bar x_n)
\end{align}
Equation~\ref{FSAD} defines the self-anti-dual property: a function remains unchanged when all inputs are complemented while the output remains uncomplemented. This property is equally valuable for inverter redistribution strategies.
For example, the 3-input majority function is a self-dual function, while the 2-input XOR is a self-anti-dual function~\cite{STLS}. For the simplicity of presentation, in this paper, we denote
\begin{align}
    f(x_1, x_2, x_3, \cdots, x_n)
\end{align}
by $f$, and denote
\begin{align}
    \bar f(\bar x_1, \bar x_2, \bar x_3, \cdots, \bar x_n)
\end{align}
by $f^d$.

\begin{theorem}\label{SDNum}
  There are $2^{2^{n-1}}$ different self-dual functions of $n$ variables~\cite{STLS}.
\end{theorem}
This exponential growth in \textbf{Theorem}~\ref{SDNum} demonstrates that the space of self-dual functions expands rapidly with input dimension, providing substantial opportunities for transformation in practical circuits with moderate cut sizes.

\begin{theorem}\label{SDComp}
  Substituting a variable in a self-dual function by another self-dual function yields a self-dual function~\cite{STLS}.
\end{theorem}
This compositional property in \textbf{Theorem}~\ref{SDComp} is crucial for our approach, as it guarantees that hierarchical transformations preserve self-duality, enabling equivalent iterative application of our redistribution method.

\subsection{And-Inverter Graph}

And-Inverter Graphs (AIGs) represent one of the most widely adopted intermediate representations for Boolean networks in modern logic synthesis tools. An AIG consists exclusively of two-input AND gates, with logical negation implemented through complemented edge attributes rather than explicit NOT gates; in this work, the terms \emph{complemented edges} and \emph{inverters} are used interchangeably. Unlike canonical forms such as ROBDD~(Reduced Ordered Binary Decision Diagrams), AIGs admit multiple structurally distinct representations for identical Boolean functions, providing flexibility for optimization. When structural hashing is applied, the AIG maintains the property that no two distinct nodes share identical fanins, ensuring compactness and facilitating efficient equivalence checking.

\subsection{Logic optimization}

Logic optimization encompasses diverse methodologies for simplifying Boolean expressions, broadly categorized into two-level and multi-level optimization techniques. Classical algorithms including Espresso~\cite{Espresso} for two-level minimization, SIS~\cite{SIS} and MVSIS~\cite{MVSIS} for multi-level synthesis, are the established foundational approaches, while contemporary state-of-the-art optimization flows typically employ sophisticated heuristic combinations, integrating techniques such as rewriting, refactoring, balancing~\cite{ClassicOpt}, and resubstitution~\cite{Mishchenko2006ScalableLS}. These methods are comprehensively implemented in the widely-used logic synthesis tool ABC~\cite{abc}. Practitioners may construct optimization recipes by strategically sequencing these operations to explore the vast solution space. These aggressive optimization strategies can yield substantial improvements in power, performance, and area (PPA) metrics for the final mapped designs.

\subsection{Cuts}
Cuts are defined on the transitive fanin of each AND node within an And-Inverter Graph. A cut comprises a boundary set of leaf nodes and a connected subgraph of internal nodes. A \textit{K-feasible} cut~\cite{FactorCuts} denotes a cut whose leaf count does not exceed a parameter \textit{K}, limiting complexity for cut enumeration and analysis.

\subsubsection{Tree/DAG nodes}
Tree nodes exhibit single fanout and form non-branching paths, while DAG~(Directed Acyclic Graph) nodes possess multiple fanouts, indicating logical sharing across distinct fanin cones~\cite{FactorCuts}.
\subsubsection{Tree cuts}
Tree cuts consist exclusively of tree nodes as internal vertices, though DAG nodes may appear as leaf inputs. These structures are synonymous with fanout-free cones, enabling isolation for local transformations.

\subsection{Level Slack}
Level slack provides a criticality metric for generic intermediate representation graphs, enabling identification of nodes on or near timing-critical paths through the comparison of required and arrival levels.

The level slack is estimated as follows:
\begin{align}\label{L_s}
    L_s = L_{rq} - L_{arr}
\end{align}
Equation~\ref{L_s} defines slack as the margin between required and arrival timing.
$L_{rq}$ is the required level, which is calculated as follows
\begin{align}\label{L_rq}
    L_{rq} = L_m - L_r
\end{align}
Equation~\ref{L_rq} computes required timing through backward traversal: The required level $L_{rq}$ at each node derives from the network's maximum logic depth $L_m$ minus the reverse topological distance $L_r$ to primary outputs.

\subsection{Choices}

The concept of choice network originates from lossless synthesis frameworks~\cite{Choices}, capitalizing on the observation that optimization sequences generate functionally equivalent yet structurally diverse nodes. These equivalent representations are captured through network snapshots across different optimization stages. Following SAT-based equivalence checking,
choice candidates are integrated into the network using linked list data structures, maintaining associations between representative and candidate nodes.  This structural diversity plays a pivotal role during technology mapping by providing alternative implementation options that can be selected to optimize quality metrics.
In the choice network, we refer to the graph after removing all the choice candidates as the primary graph. \looseness=-1

\section{Inverter Redistribution}\label{InvR}
Our method employs cut-based analysis on AIG representations to identify self-dual and self-anti-dual subfunctions, enabling targeted inverter redistribution.
The redistribution is strategically positioned after comprehensive combinational optimization but preceding technology mapping, ensuring optimal structural conditions for transformation.

Overall, the approach traverses the AIG  to identify self-dual and self-anti-dual functions (with respect to a local cut) and then selectively applies transformations following a certain policy to ensure targeted redistribution of inverters. In this section, we will first present the method of detecting self-duality and self-anti-duality in Section~\ref{subsec:detection}, followed by a brief discussion of the transformation in Section~\ref{subsec:transformation}. Section~\ref{subsec:node-select} and \ref{subsec:cut-select} discuss about what nodes and cuts are targeted during the traversal, ensuring an efficient and sound implementation of the algorithm. Finally, Section~\ref{subsec:redist-policy} presents the policy of redistribution. \looseness=-1

\subsection{Self-Dual/Self-Anti-Dual Function Detection}
\label{subsec:detection}

During traversal of all AND nodes within the current AIG, we employ exact simulation techniques to evaluate functional properties.
Exact simulation computes the truth table of the current node $n_c$ with respect to its cut leaf set $L_n$, producing the truth vector $\texttt{uTruth}$ that corresponds to a function $f$.
Subsequently, we complement all leaf inputs and perform a second simulation round, generating $\texttt{uTruthFlipped}$ which represents $\bar{f^d}$ as previously mentioned in Section~\ref{SDSAD}.
By comparing the phase indicators $\texttt{fPhaseOri}$ and $\texttt{fPhaseFlipped}$ against the truth table vectors $\texttt{uTruth}$ and $\texttt{uTruthFlipped}$, we determine whether the function exhibits self-dual or self-anti-dual properties.
The algorithm is shown in Algorithm~\ref{alg:SDSADPseudocode}.
A masking vector $\texttt{uMask}$ is employed to identify valid bits within the simulated truth table, ensuring accurate comparison for cut sizes smaller than the maximum supported input count. We only use cuts that have less than or equal to 5 leaves, with \texttt{uMask} being one of the following $\{\text{0x3},\text{0xF}, \text{0xFF}, \text{0xFFFF}, \text{0xFFFFFFFF}\}$, for 1- to 5-feasible cuts, respectively.

\begin{algorithm}[h]
\caption{Self-Dual / Self-Anti-Dual Detection}
\label{alg:SDSADPseudocode}
\small
\begin{algorithmic}[1]
\Function{DetectSDOrSAD}{$uTruth$, $uTruthFlipped$, $uMask$, $fPhaseOri$, $fPhaseFlipped$}
    \State $isComp \leftarrow (uTruth = \neg uTruthFlipped) \wedge uMask$
    \State $truthEq \leftarrow (uTruth = uTruthFlipped)$
    \State $phaseEq \leftarrow (fPhaseOri = fPhaseFlipped)$
    \State $phaseXor \leftarrow (fPhaseOri \oplus fPhaseFlipped = 1)$
    \If{$truthEq \land phaseXor$}
        \Return \text{Self-Dual}
    \ElsIf{$isComp \land phaseEq$}
        \Return \text{Self-Dual}
    \ElsIf{$truthEq \land phaseEq$}
        \Return \text{Self-Anti-Dual}
    \ElsIf{$isComp \land phaseXor$}
        \Return \text{Self-Anti-Dual}
    \Else
    \Statex
        \Return \text{Not Self-Dual/Anti-Dual}
    \EndIf
\EndFunction
\end{algorithmic}
\end{algorithm}

\subsection{AIG Transformation}
\label{subsec:transformation}

Once self-dual or self-anti-dual properties are confirmed, the  transformation could take place (if it is permitted by the policy that will be presented in Section~\ref{subsec:redist-policy}).
The transformation involves modification of fanin and fanout edge complementation attributes.
Specifically, when a node is identified as implementing a self-anti-dual function relative to a specific cut, the transformation will complement the edges at the cut leaves.
Note that the complementation is applied only to edges feeding internal nodes of the cut boundary, which is  essential for maintaining logical equivalence during transformation.
For  self-dual functions, it is also necessary to complement fanout edges of the root node.

\subsection{Node Selection in AIG Traversal}
\label{subsec:node-select}

The above transformation targets two distinct node categories in the AIG: (1) DAG nodes and tree nodes on the primary graph lacking choice structures, and (2) choice candidate nodes within the choices network.
Given the demonstrated efficacy of choice networks in improving mapping quality, we extend our redistribution methodology to include choice candidates, maximizing optimization opportunities. Nodes with choices are not necessarily DAG nodes; they may also be tree nodes. During traversal, these choice representatives are ignored, while their candidates are considered.
Our empirical observation suggests that tree nodes on the primary graph associated with choice candidates exhibit optimization resistance following intensive optimization flows, implying that the primary structure has already reached a near-optimal configuration and should be preserved.

\subsection{Cut Selection in AIG Traversal}
\label{subsec:cut-select}

Our methodology proceeds through systematic iteration over tree cuts. This design ensures that transformations operate within well-defined structural boundaries.
By collecting tree cuts exclusively, we guarantee that all identified cuts reside within their corresponding Maximum Fanout Free Cones~(MFFCs), maintaining structural isolation.
Tree cuts exhibit mutual isolation, equivalent to fanout-free cones, enabling safe complemented attribute modifications without affecting other nodes' functionality. This structural independence ensures that self-anti-dual transformations preserve equivalence.
For self-dual transformations, \textbf{Theorem}~\ref{SDComp} guarantees that even when subsequent transformations affect previously modified edges within new cut windows, logical equivalence is maintained due to compositional closure properties.
We prohibit DAG cuts due to their structural complexity and interaction patterns, which can introduce equivalence preservation challenges.

\begin{figure}[h]
    \centering
    \includegraphics[scale=0.18]{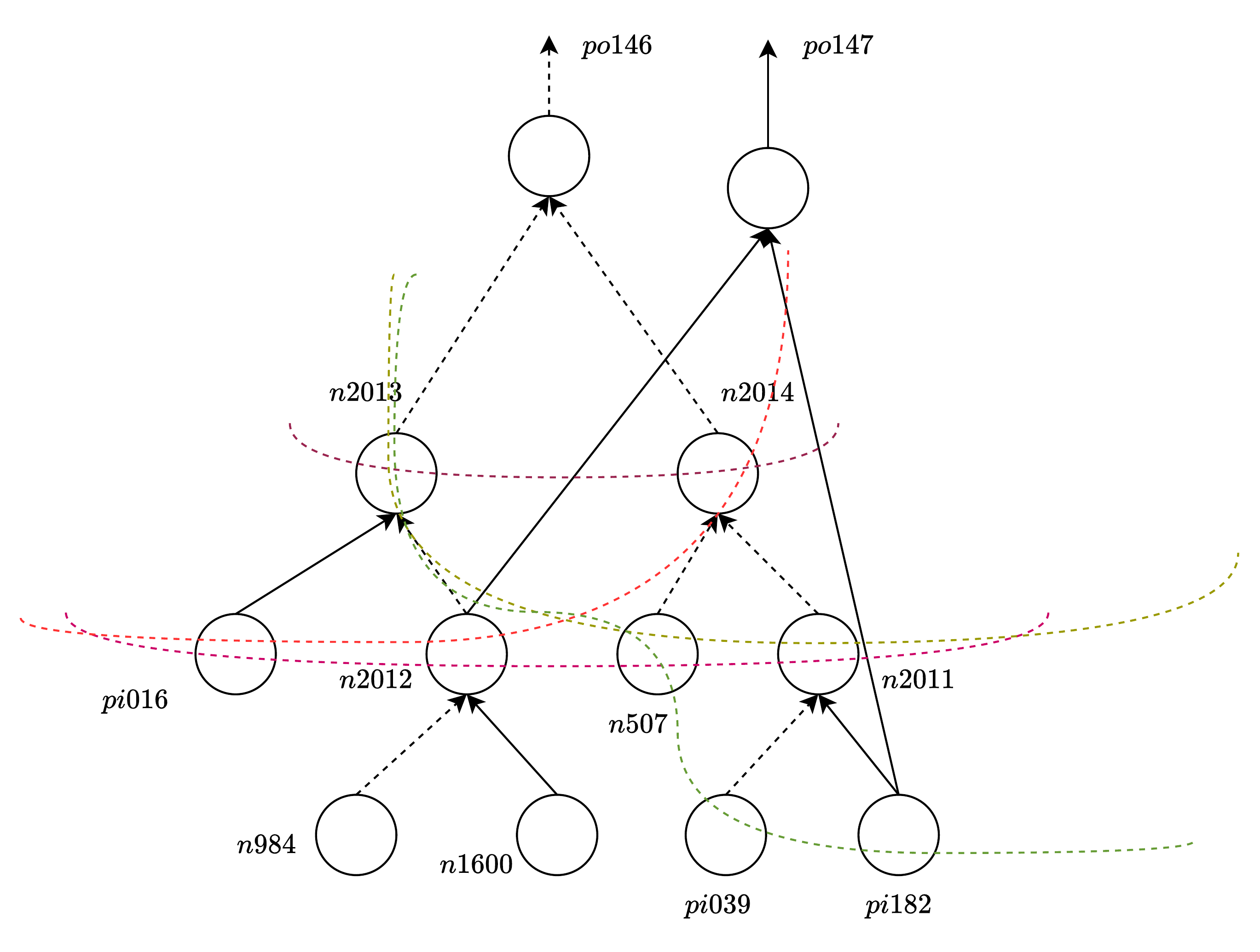}
    \caption{Among the 4-feasible cuts of node $po146$, there are 5 tree cuts as the dashed curves show. $\{pi016, n984, n1600, n2014\}$ is excluded even though it is a 4-feasible cut, since node $n2012$ is a DAG node.
    }
    \label{fig:TreeCut}
\end{figure}

\begin{figure}[h]
    \centering
    \includegraphics[scale=0.44]{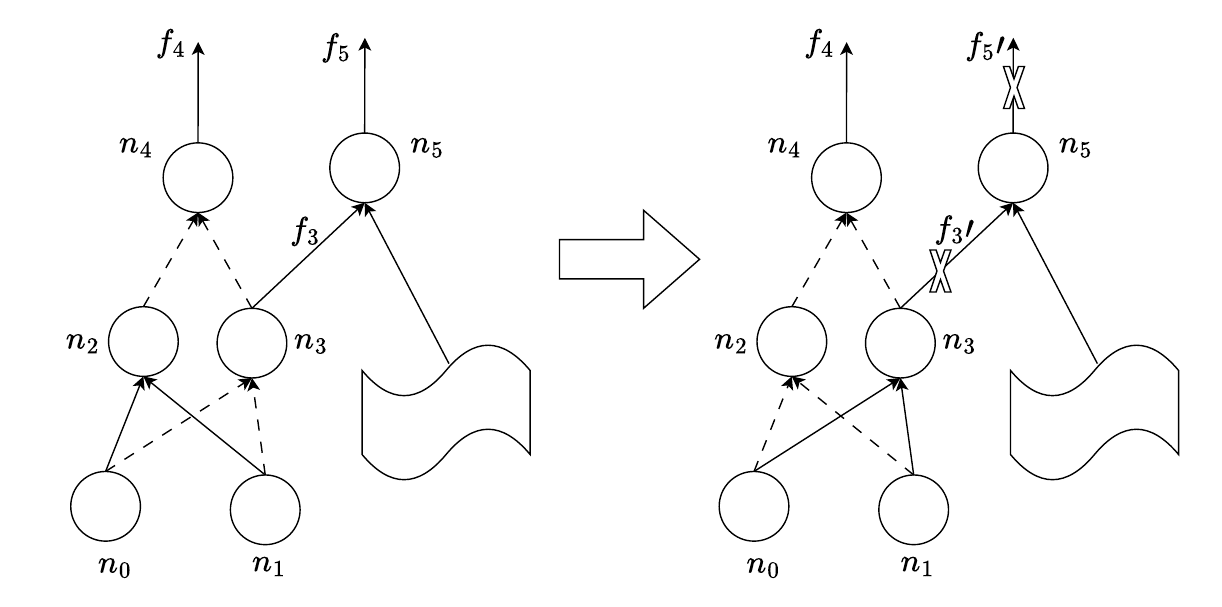}
    \caption{ Dashed lines are complemented edges. Solid lines are regular (uncomplemented) edges. $n_4$ is representing an XOR in AIG, and there is a valid but different implementation of XOR on the right side, which could be achieved by inverter redistribution, but since $n_3$ has multiple fanouts, even though $n_4$ remains equivalent, $f_3'$ is not equivalent to $f_3$, which leads to $f_5'$ being not equal to $f_5$. Therefore, the whole design is not equivalent to the original one. This demonstrates it is not safe to use DAG cuts in inverter redistribution.
    }
    \label{fig:nonEq}
\end{figure}
Figure~\ref{fig:TreeCut} shows an example of a tree cut set. The tree-cut-based method is safe and independent, even changing the internal edges on leaves will not affect the equivalence results. In contrast, Figure~\ref{fig:nonEq} demonstrates an example where utilizing DAG cuts would lead to equivalence failure after redistribution.

\subsection{Redistribution Policy}
\label{subsec:redist-policy}
\subsubsection{Critical and Non-Critical Path Redistribution}
For nodes identified as critical through level slack analysis, we restrict equivalent inverter redistribution to cases where the complemented edge count on the critical path exceeds the regular edge count.
This selective application ensures net reduction of inverter count on timing-critical paths, directly targeting delay improvement.
For nodes that are not on critical paths, we permit redistribution when the total inverter count meets or exceeds half of all edges, enabling more aggressive structural modifications.

\subsubsection{Zero-Gain Redistribution on Non-Critical Paths}
The XOR operation exhibits self-anti-dual properties, as previously discussed in Section~\ref{SDSAD}. Tree-structured XORs are defined as XOR configurations within an And-Inverter Graph (AIG) that contain no internal fanouts. In most scenarios, zero-gain inverter redistribution applied to non-critical paths involves tree-structured two-input XOR functions. Due to the inherent symmetry of the XOR operation, such redistribution on non-critical paths does not yield delay improvements. However, it creates opportunities for enhanced inverter absorption during subsequent technology mapping.

This effect is particularly evident in the cases \textit{voter}, \textit{hyp}, \textit{sqrt}, and \textit{square}, where the AIG-level complemented edge count remains unchanged, yet a substantial absorption of \textbf{INV\_X1} cells could occur after mapping, leading to notable delay reductions. Although zero-gain inverter redistribution on non-critical paths preserves the local inverter ratio at 0.5---hence the term ``zero-gain"---it strategically relocates inverters within the network. This reduces the capacitive load on critical-path gates and enables more efficient mapping decisions, such as the preferential use of XNOR2 cells (44.4 ps) over XOR2 cells (53.5 ps), and NAND2 cells (21 ps) instead of slower AOI21 cells (43 ps) when mapping on the \textit{Nangate45} technology library.

The proposed one-way algorithm applies these irreversible transformations exclusively to non-critical cuts, thereby avoiding any adverse impact on critical paths. This targeted approach yields cumulative delay improvements of 3--4\% in arithmetic circuits. A similar method applied in previous work on technology mapper \textit{\&nf} in ABC has the term of ``pin-swap" or ``pin-permutation." However, the prior method is done on-the-fly during mapping evaluation, while here we do it on the non-critical tree-XOR in AIG, lower the chance of suboptimal delay.

\section{Experiments}\label{EXP}
\subsection{Experimental Setup}

Experiments are conducted on a machine equipped with an Apple M4 Pro processor and 24 GB of memory, running Tahoe 26.2. The proposed method is implemented within the ABC framework~\cite{abc}, operating entirely on AIG representations. Functional correctness after inverter redistribution is verified using the combinational equivalence checking command \texttt{cec} in ABC. The redistribution process is triggered by the command \texttt{rd\_inv -s -c}. We restrict our analysis to 5-feasible cuts, as a function with at most five variables can be represented by a single 32-bit unsigned integer for truth table simulation. As stated in \textbf{Theorem}~\ref{SDNum}, the space of self-dual functions with five variables contains $2^{2^{4}}$ distinct functions, which is sufficient to demonstrate the effectiveness of our approach.

\subsection{EPFL Combinational Benchmark}

The EPFL combinational benchmark suite~\cite{EPFLCombBench} is used as the test vehicle, covering both random control logic and arithmetic logic. The large \textit{More than ten million gates}~(MTM) benchmarks are excluded due to their scale. The following synthesis flows are compared:
\begin{itemize}
    \item OptFlow: $(\text{st; \&get; \&if -g; \&dch; \&put})$
    \item Baseline: $(\text{r2rs})^8(\text{OptFlow})^8(\text{map; topo; stime})$
    \item Ours: $(\text{r2rs})^8(\text{OptFlow})^8(\text{rd\_inv -s -c})(\text{map; topo; stime})$
\end{itemize}
In OptFlow, \texttt{st} performs structural hashing~\cite{ClassicOpt}; \texttt{\&get} and \texttt{\&put} are ABC9-specific structural transformation commands; \texttt{\&if -g} performs SOP-based balancing~\cite{sopb}, which achieves stronger algebraic balancing than the standard \texttt{balance}~\cite{ClassicOpt} command; and \texttt{\&dch}~\cite{Choices} constructs a choice network after extensive optimization. The \texttt{r2rs} recipe applies level-constrained rewriting and resubstitution that disallow any logic level increase, serving as a classical recipe for delay-oriented optimization. The \texttt{map} command performs standard cell mapping using a genlib derived from the \textit{Nangate45} technology library, producing representative cells without cell sizing. The \texttt{topo} and \texttt{stime} commands perform topological reordering and static timing analysis, respectively. Note that choices constructed in earlier iterations are removed by structural hashing as dangling nodes; only choices from the final optimization round are preserved for mapping.

The superscripts in \textbf{Baseline} and \textbf{Ours} denote the number of repeated iterations. Since inverter redistribution does not modify AND nodes, the level slack remains unchanged after each application of \texttt{rd\_inv -s -c}, and only a single round is needed. For the \textit{hyp} benchmark, which has a considerably larger scale, \texttt{OptFlow} is doubled to 16 iterations in both \textbf{Baseline} and \textbf{Ours} to ensure sufficient optimization; for all other benchmarks, 8 iterations suffice.

Table~\ref{rmiCuts} reports the synthesis results. In the \textbf{InvNum} column, the format \textbf{A/B} denotes the number of complemented edges in the AIG (A) and the number of \textbf{INV\_X1} cells in the mapped network (B), respectively. Columns \textbf{Gates}, \textbf{Delay~(ps)}, and \textbf{Area} report post-mapping metrics.

\subsection{Analysis of the result}

As shown in Table~\ref{rmiCuts}, \textbf{Ours} achieves no delay degradation on any benchmark compared to \textbf{Baseline}, with delay reductions observed primarily on arithmetic circuits. In contrast, random control logic benchmarks show no change, as the inverter distribution in these circuits remains unaffected by our method. We attribute this to the highly structured nature of arithmetic logic, which contains a large number of self-dual or self-anti-dual subfunctions amenable to our transformation~\cite{DC_UG_Ch3_2025}.

An interesting observation from the \textbf{InvNum} column is that inverter redistribution, rather than inverter removal, is the primary contributor to the improved results: the AIG-level complemented edge count remains unchanged in most cases, yet the mapped \textbf{INV\_X1} count decreases. This confirms that strategic inverter relocation prior to mapping enables more efficient cell selection during technology mapping. Gate count and area generally decrease as well, though area tradeoffs occur in cases such as \textit{sqrt}, where delay improvement takes priority.

Among the arithmetic benchmarks, \textit{sqrt} achieves the largest delay reduction of 3.86\%, followed by \textit{voter} (1.09\%), \textit{log2} (0.85\%), \textit{div} (0.10\%), \textit{square} (1.60\%), and \textit{hyp}$^{16}$ (1.46\%). The \textit{multiplier} and \textit{sin} benchmarks show modest improvements of 0.77\% and 0.04\%, respectively.

\begin{figure}[h]
    \centering
    \includegraphics[scale=0.61]{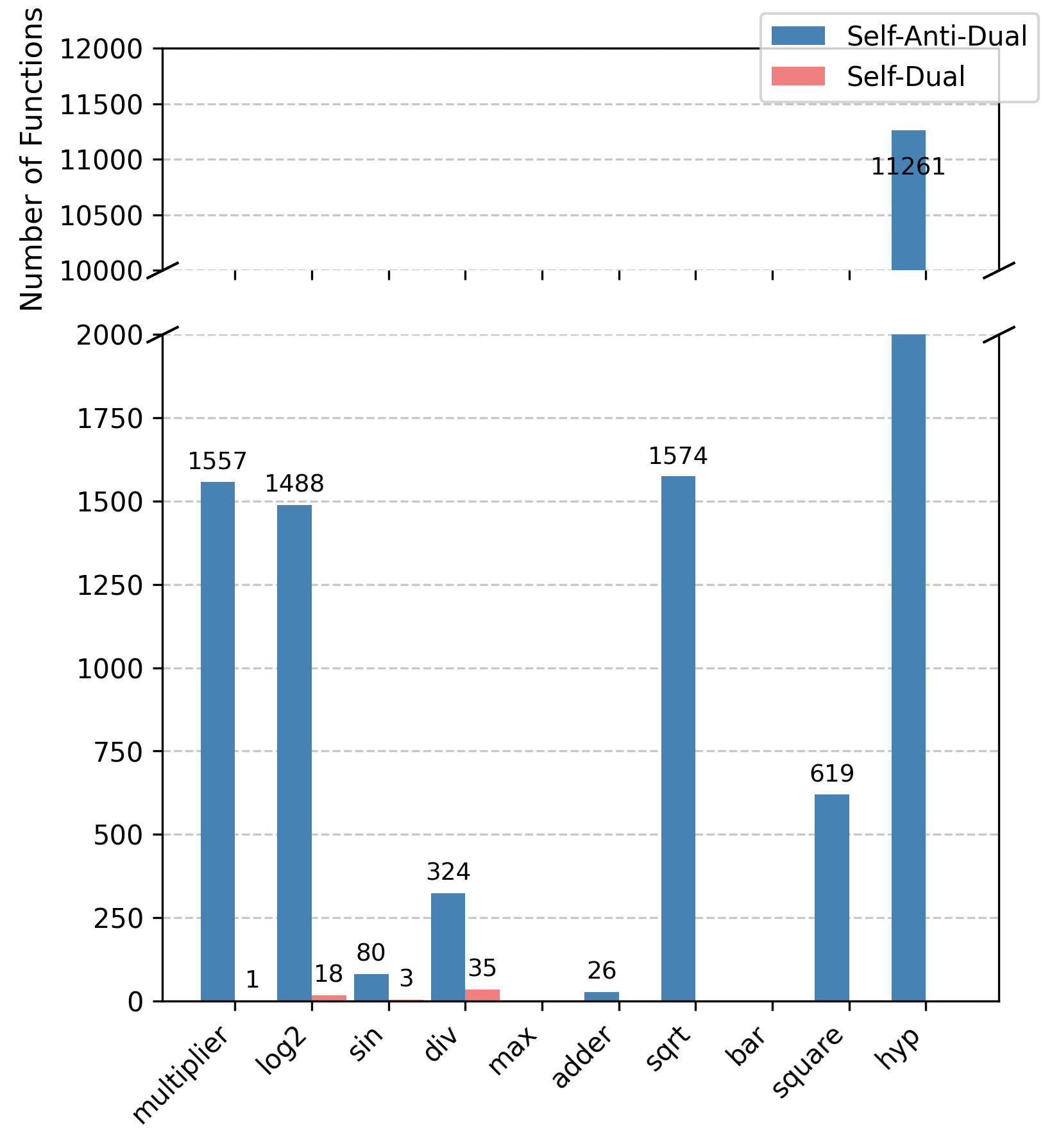}
    \caption{Number of self-dual and self-anti-dual subfunctions identified in each arithmetic benchmark.}
    \label{fig:CountSDSAD}
\end{figure}

\begin{figure}[h]
    \centering
    \includegraphics[scale=0.32]{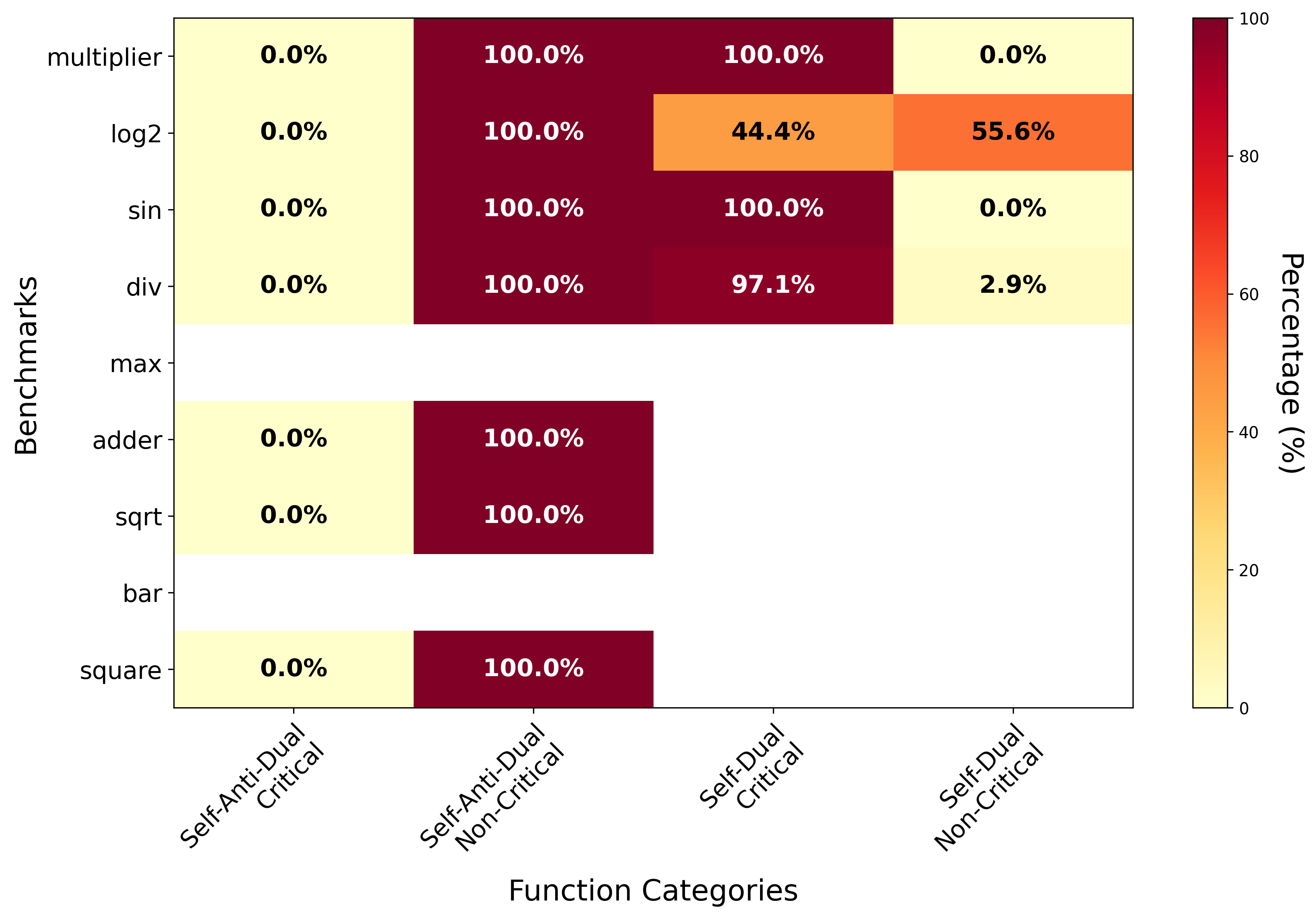}
    \caption{Critical vs.\ non-critical distribution of identified self-dual and self-anti-dual subfunctions.}
    \label{fig:hmCriRatio}
\end{figure}

Figure~\ref{fig:CountSDSAD} shows the number of self-dual and self-anti-dual subfunctions identified in each arithmetic benchmark. Self-anti-dual functions---primarily tree-structured XORs---vastly outnumber self-dual functions, indicating that self-anti-dual redistribution is the dominant contributor to the observed improvements. Figure~\ref{fig:hmCriRatio} presents the critical versus non-critical distribution of identified functions, revealing that self-dual functions appear exclusively on critical paths, while self-anti-dual functions are found exclusively on non-critical paths. This distribution aligns with our evaluation strategy: self-dual redistribution directly reduces inverters on critical paths, while self-anti-dual redistribution on non-critical paths creates favorable conditions for subsequent mapping, as discussed in Section~\ref{InvR}.

\begin{table*}[htbp]
\setlength{\tabcolsep}{5.0 pt}
\centering
\caption{Comparison of Synthesis Results: \texttt{Baseline} vs \texttt{Ours}}
\label{rmiCuts}
\begin{tabular}{@{}lcccccccccccc@{}}
\toprule
\multirow{2}{*}{\textbf{Benchmark}} & \multicolumn{4}{c}{\textbf{Baseline}} & \multicolumn{4}{c}{\textbf{Ours}} & \multicolumn{4}{c}{\textbf{Difference (Ours - Baseline)}} \\
\cmidrule(lr){2-5} \cmidrule(lr){6-9} \cmidrule(lr){10-13}
& \textbf{InvNum} & \textbf{Gates}  & \textbf{Delay (ps)} & \textbf{Area} & \textbf{InvNum} & \textbf{Gates} &  \textbf{Delay (ps)} & \textbf{Area} & \textbf{InvNum} & \textbf{Gates} & \textbf{Delay (ps)} & \textbf{Area} \\
\midrule
cavlc & 693/34 & 405 & 303.44 & 391.55 & 693/34 & 405 & 303.44 & 391.55 & 0/0 & 0 & 0.00 & 0.00 \\
priority & 464/85 & 378 & 265.95 & 343.41 & 464/85 & 378 & 265.95 & 343.41 & 0/0 & 0 & 0.00 & 0.00 \\
ctrl & 96/13 & 93 & 159.55 & 86.45 & 96/13 & 93 & 159.55 & 86.45 & 0/0 & 0 & 0.00 & 0.00 \\
mem\_ctrl & 40464/3226 & 24661 & 3856.61 & 23595.00 & 40464/3226 & 24662 & 3856.61 & 23595.26 & 0/0 & +1 & 0.00 & +0.26 \\
dec & 16/8 & 300 & 190.63 & 254.03 & 16/8 & 300 & 190.63 & 254.03 & 0/0 & 0 & 0.00 & 0.00 \\
int2float & 206/15 & 145 & 199.61 & 139.38 & 206/15 & 145 & 199.61 & 139.38 & 0/0 & 0 & 0.00 & 0.00 \\
\rowcolor{green!20} voter & 12605/1265 & 11850 & 1276.07 & 11423.64 & 12605/1255 & 11828 & 1262.14 & 11407.14 & 0/-10 & -22 & -13.93 & -16.50 \\
router & 140/23 & 149 & 246.68 & 115.71 & 140/23 & 149 & 246.68 & 115.71 & 0/0 & 0 & 0.00 & 0.00 \\
i2c & 1128/121 & 752 & 388.13 & 709.42 & 1128/121 & 752 & 388.13 & 709.42 & 0/0 & 0 & 0.00 & 0.00 \\
arbiter & 4799/661 & 4295 & 407.35 & 4044.00 & 4799/661 & 4295 & 407.35 & 4044.00 & 0/0 & 0 & 0.00 & 0.00 \\
\midrule
\rowcolor{green!20} $\text{hyp}^{16}$ & 326549/36502 & 290616 & 715756.38 & 279980.16 & 326549/34543 & 290255 & 705318.44 & 279722.16 & 0/-1959 & -361 & -10437.94 & -258.00 \\
\rowcolor{green!20}multiplier & 35502/2651 & 27646 & 3481.72 & 28466.26 & 35501/2631 & 27471 & 3454.88 & 28313.57 & -1/-20 & -175 & -26.84 & -152.69 \\
\rowcolor{green!20}log2 & 37896/2818 & 32917 & 8715.05 & 33259.84 & 37886/2835 & 32816 & 8641.18 & 33143.60 & -10/+17 & -101 & -73.87 & -116.24 \\
\rowcolor{green!20}sin & 6667/639 & 6595 & 3842.20 & 6596.80 & 6664/673 & 6591 & 3840.65 & 6550.52 & -3/+34 & -4 & -1.55 & -46.28 \\
\rowcolor{green!20} div & 64556/6932 & 59294 & 61347.12 & 57626.24 & 64518/6940 & 59208 & 61285.65 & 57562.13 & -38/+8 & -86 & -61.47 & -64.11 \\
max & 3538/607 & 3311 & 4359.38 & 3172.05 & 3538/607 & 3311 & 4359.38 & 3172.05 & 0/0 & 0 & 0.00 & 0.00 \\
adder & 1841/269 & 1499 & 528.68 & 1402.62 & 1841/269 & 1499 & 528.68 & 1402.62 & 0/0 & 0 & 0.00 & 0.00 \\
\rowcolor{green!20} sqrt & 33348/5311 & 32596 & 136988.27 & 31388.80 & 33348/4882 & 33025 & 131698.45 & 31417.79 & 0/-429 & +429 & -5289.82 & +28.99 \\
bar & 3852/135 & 2710 & 1914.18 & 2854.98 & 3852/135 & 2710 & 1914.18 & 2854.98 & 0/0 & 0 & 0.00 & 0.00 \\
\rowcolor{green!20} square & 22153/2259 & 20513 & 1104.39 & 19898.93 & 22153/2244 & 20484 & 1086.70 & 19879.78 & 0/-15 & -29 & -17.69 & -19.15 \\
\midrule
\textbf{Geo Mean} & \textbf{3605.5/399.7} & \textbf{3511.0} & \textbf{1885.1} & \textbf{3337.7} & \textbf{3605.3/397.6} & \textbf{3510.5} & \textbf{1875.9} & \textbf{3334.5} & \textbf{---} & \textbf{---} & \textbf{---} & \textbf{---} \\
\bottomrule
\end{tabular}
\end{table*}

\section{Conclusion}\label{CON}
Arithmetic circuits exhibit higher optimization complexity than random control logic under classical synthesis approaches. This work exploits self-dual and self-anti-dual properties to enable pre-mapping inverter redistribution, yielding measurable delay improvements on arithmetic benchmarks. For random control logic, our method maintains quality without degradation, representing a safe and incremental enhancement applicable to broad synthesis workflows.

\bibliographystyle{ieeetr}
\bibliography{reference}

\begin{thebibliography}{10}

\bibitem{Orchestrate}
Y.~Li, M.~Liu, H.~Ren, A.~Mishchenko, and C.~Yu, ``Dag-aware synthesis
  orchestration,'' {\em IEEE Transactions on Computer-Aided Design of
  Integrated Circuits and Systems}, vol.~43, no.~12, pp.~4666--4675, 2024.

\bibitem{ExactDSyn}
L.~Amar\'{u}, M.~Soeken, P.~Vuillod, J.~Luo, A.~Mishchenko, P.-E. Gaillardon,
  J.~Olson, R.~Brayton, and G.~De~Micheli, ``Enabling exact delay synthesis,''
  in {\em Proceedings of the 36th International Conference on Computer-Aided
  Design}, ICCAD '17, p.~352–359, IEEE Press, 2017.

\bibitem{speedup}
A.~Mishchenko, R.~Brayton, and S.~Jang, ``Global delay optimization using
  structural choices,'' in {\em Proceedings of the 18th Annual ACM/SIGDA
  International Symposium on Field Programmable Gate Arrays}, FPGA '10, (New
  York, NY, USA), p.~181–184, Association for Computing Machinery, 2010.

\bibitem{Hazard}
E.~B. Eichelberger, ``Hazard detection in combinational and sequential
  switching circuits,'' {\em IBM Journal of Research and Development}, vol.~9,
  no.~2, pp.~90--99, 1965.

\bibitem{10SimpleRules}
B.~C. Haller, ``Ten simple rules for reporting a bug,'' {\em PLOS Computational
  Biology}, vol.~18, pp.~1--8, 10 2022.

\bibitem{STLS}
T.~Sasao, {\em Switching Theory for Logic Synthesis}.
\newblock USA: Kluwer Academic Publishers, 1st~ed., 1999.

\bibitem{Espresso}
R.~K. Brayton, A.~L. Sangiovanni-Vincentelli, C.~T. McMullen, and G.~D.
  Hachtel, {\em Logic Minimization Algorithms for VLSI Synthesis}.
\newblock USA: Kluwer Academic Publishers, 1984.

\bibitem{SIS}
E.~Sentovich, K.~Singh, L.~Lavagno, C.~Moon, R.~Murgai, A.~Saldanha, H.~Savoj,
  P.~Stephan, R.~K. Brayton, and A.~L. Sangiovanni-Vincentelli, ``Sis: A system
  for sequential circuit synthesis,'' Tech. Rep. UCB/ERL M92/41, May 1992.

\bibitem{MVSIS}
R.~Brayton, M.~Gao, J.-H. Jiang, Y.~Jiang, Y.~Li, A.~Mishchenko, S.~Sinha, and
  T.~Villa, ``Optimization of multi-valued multi-level networks,'' in {\em
  Proceedings of the 32nd International Symposium on Multiple-Valued Logic},
  ISMVL '02, (USA), p.~168, IEEE Computer Society, 2002.

\bibitem{ClassicOpt}
A.~Mishchenko, S.~Chatterjee, and R.~Brayton, ``Dag-aware aig rewriting a fresh
  look at combinational logic synthesis,'' in {\em Proceedings of the 43rd
  Annual Design Automation Conference}, DAC '06, (New York, NY, USA),
  p.~532–535, Association for Computing Machinery, 2006.

\bibitem{Mishchenko2006ScalableLS}
A.~Mishchenko and R.~K. Brayton, ``Scalable logic synthesis using a simple
  circuit structure,'' 2006.

\bibitem{abc}
R.~Brayton and A.~Mishchenko, ``Abc: an academic industrial-strength
  verification tool,'' in {\em Proceedings of the 22nd International Conference
  on Computer Aided Verification}, CAV'10, (Berlin, Heidelberg), p.~24–40,
  Springer-Verlag, 2010.

\bibitem{FactorCuts}
S.~Chatterjee, A.~Mishchenko, and R.~Brayton, ``Factor cuts,'' in {\em
  Proceedings of the 2006 IEEE/ACM International Conference on Computer-Aided
  Design}, ICCAD '06, (New York, NY, USA), p.~143–150, Association for
  Computing Machinery, 2006.

\bibitem{Choices}
S.~Chatterjee, A.~Mishchenko, R.~Brayton, X.~Wang, and T.~Kam, ``Reducing
  structural bias in technology mapping,'' in {\em ICCAD-2005. IEEE/ACM
  International Conference on Computer-Aided Design, 2005.}, pp.~519--526,
  2005.

\bibitem{EPFLCombBench}
L.~Amarù, P.-E. Gaillardon, and G.~De~Micheli, ``The epfl combinational
  benchmark suite,'' 2015.

\bibitem{sopb}
A.~Mishchenko, R.~Brayton, S.~Jang, and V.~Kravets, ``Delay optimization using
  sop balancing,'' in {\em Proceedings of the International Conference on
  Computer-Aided Design}, ICCAD '11, p.~375–382, IEEE Press, 2011.

\bibitem{DC_UG_Ch3_2025}
{Synopsys, Inc.}, {\em Design Compiler User Guide}.
\newblock Synopsys, Inc., Mountain View, CA, USA, Jan. 2025.
\newblock Chapter 3.

\end{thebibliography}

\vspace{12pt}
\end{document}